\documentclass[a4paper,11pt]{article}
\usepackage[utf8]{inputenc}

\usepackage{amsfonts}
\usepackage{amssymb}
\usepackage{graphicx}
\usepackage{amsmath}
\usepackage{enumerate}
\usepackage{color}
 \usepackage{multirow}
\usepackage{float}

\RequirePackage{mathrsfs} \RequirePackage[sc]{mathpazo}
\RequirePackage{wasysym} \RequirePackage{setspace}\textheight=650pt
\title{ {\bf On Thermodynamics of AdS Black Holes in M-Theory}}
\author{A. Belhaj$^{1,2}$, M. Chabab$^2$, H. EL Moumni$^2$, K. Masmar$^2$, M.  B. Sedra$^{3}$\\
\\
{\small $^{1}$D\'epartement de Physique, LIRST, Facult\'e
Polydisciplinaire, Universit\'e Sultan
 Moulay Slimane } \\  {\small  B\'eni Mellal,  Morocco. } \\
{\small $^{2}$High Energy Physics and Astrophysics Laboratory, FSSM,
 \small Cadi Ayyad University, Marrakesh, Morocco.
} \\
{\small $^{3}$  D\'{e}partement de Physique, SIMO, Facult\'{e} des
Sciences, Universit\'{e} Ibn Tofail,
 K\'{e}nitra, Morocco.} 
 }
\date{}

\begin{document} \maketitle
\begin{abstract}
Motivated by a  recent work on asymptotically Ad$S_4$
black holes in M-theory, we investigate   the thermodynamics and
thermodynamical geometry of AdS black holes from M2 and M5-branes.
Concretely, we consider     AdS black holes in $AdS_{p+2}\times
S^{11-p-2}$, where $p=2,5$ by interpreting  the number of M2 (and
M5-branes) as a thermodynamical variable.  We study  the
corresponding phase transition to examine   their stabilities by
calculating and discussing  various thermodynamical quantities
including the chemical potential. Then, we compute  the
thermodynamical curvatures from the  Quevedo metric for M2 and
M5-branes geometries to reconsider  the stability of such black
objects. The Quevedo metric singularities recover similar stability
results provided by the phase transition program.
\end{abstract}
\newpage
\section{Introduction}

Recently, an increasing interest has been devoted to the study of
the black  hole physics   in  connection with many subjects
including string theory and  famous thermodynamical models.   The
interest has been explored to develop    deeper   relationships
between the gravity theories and the thermodynamical physics using
anti-De Sitter geometries.  In this issue, laws of black holes  can
be identified with laws of thermodynamics \cite{30,witten,4,5,50}.
More precisely, the phase transition and the  critical phenomena for
various AdS black holes have been  extensively  investigated using
different approachs \cite{6,a1,Dolan,Cliff,15Dolan}. In this way,
certain equations of state,  describing   rotating black holes, have
been identified with some known thermodynamical  ones. In
particular,  it has been remarked serious efforts discussing  the
behavior of the Gibbs free energy in the fixed charge ensemble. This
program  has led to a nice interplay between the behavior of the AdS
black hole systems and the Van der Waals
fluids\cite{our,our1,our2,hasan,KM,ourx,oury,Zhao}. In fact, it has
been shown that  P-V criticality, Gibbs free energy, first order
phase transition and the behavior near the critical points can be
associated with   the liquid-gas systems.

More recently, a special focus    has been put  on the
thermodynamics and thermodynamical geometry for a fivedimensional
AdS black hole in type IIB superstring background known by
$AdS_5\times S^5$ \cite{Zhang,Zhang1,jhep45}. It is recalled that
this geometry has been studied in many  places in connection with
AdS/CFT correspondence, providing a nice equivalence between
gravitational theories  in d-dimensional AdS  geometries and
conformal field theories (CFT) in a (d-1)-dimensional boundary of
such  AdS spaces\cite{Maldacena:1997re,ads,ads1,ads2}. In such black
hole activities,
 the number of colors has been interpreted  as  a thermodynamical
 variable. In particular, the
thermodynamic properties of  black holes  in $AdS_5\times S^5$  have
been investigated   by    considering  the cosmological constant in
the bulk as the number of colors.  In fact, many thermodynamical
quantities have been computed to discuss the stability behaviors of
such black holes.

Motivated by these activities and a  recent work   on asymptotically
Ad$S_4$ black holes in M-theory \cite{m0,m1,m2,m3}, we investigate
the thermodynamics and thermodynamical geometry of AdS black holes
from the physics of M2 and M5-branes.  Concretely, we study  AdS
black holes in $AdS_{p+2}\times S^{11-p-2}$, where $p=2,5$ by
viewing the number of M2 and M5-branes  as a thermodynamical
variable.  To discuss the stability  of  such solutions,  we examine
first  the corresponding phase transition by  computing  the
relevant quantities including  the chemical potential.  Then, we
calculate the thermodynamical curvatures from the  Quevedo metric
for M2 and M5-brane geometries to reconsider the study of the
stability.

The paper is organized as follows.   We discuss  thermodynamic
properties and stability of  the  black holes in  $AdS_{p+2}\times
S^{11-p-2}$, where $p=2,5$ by viewing the number of M2 and M5-branes
as a thermodynamical variable in section 2 and 3. Similar results
which have been recovered  using thermodynamical curvature
calculations, associated with the Quevedo metric,   are presented in
section 4. The last section is devoted to conclusion.

\section{ Thermodynamics of black holes in  $AdS_4\times S^7$  space}

In this section, we investigate  the phase transition of  the AdS
black holes in M-theory in the presence of solitonic objects. It is
recalled that, at lower energy, M-theory describes   an eleven
dimensional supergravity. This theory, which was proposed by Witten,
can produce some non perturbative limits  of superstring models
after its compactification  on  particular geometries
\cite{Witten:1997sc}.

It has been shown  that  the corresponding  eleven  supergravity
involves  a cubic $R^4$ one-loop UV divergence \cite{a54} which  has
been   obtained using a specific cutoff motivated by string theory
\cite{a55,a56}.   This calculation  gives  the following correction
of the Einstein action
\begin{equation}
I=-\int d^{11} x \sqrt{g}\left(\frac{1}{2 \kappa_{11}} R + \frac{1}{\kappa_{11}^{2/3}} \zeta W+\cdots\right),
\end{equation}
with $\kappa_{11}$ is related to the Planck length by $\kappa_{11}^2= 2^4\pi^5 \ell_{11}^2$,
 $\zeta=\frac{2\pi^2}{3}$.  $W$ can be  given  in terms   of the Ricci tensor as follows $W\sim RRRR $
 \cite{a,a36}.  Roughly speaking,  M-theory  contains two fundamental objets called M2 and M5 branes coupled in eleven dimension to 3
   and 6 forms  respectively. The
near horizon of such  black objects is defined by the product of AdS
spaces and spheres
\begin{equation}
Ads_{p+2}\times S^{11-p-2}, \qquad p=2,5
\end{equation}

 To start,
let us consider the case of  M2-brane.  The corresponding geometry
is $AdS_4\times S^7$. In such a geometric background, the line
element of the black M2-brane metric  is given by \cite{a58,a}
\begin{equation}\label{ds4}
ds^2= \frac{r^4}{L^4}\left(-f dt ^2+\sum_{i=1}^2 dx_i^2 \right)+\frac{L^2}{r^2}f^{-1}dr^2+ L^2 d\Omega_7^2,
\end{equation}
where $d\Omega_7^2$ is the metric of seven-dimensional sphere with
unit radius.  In this solution, the metric function reads as follows
\begin{equation}\label{f4}
f=1-\frac{m}{r}+\frac{r^2}{L^2},
\end{equation}
where $L$ is the AdS radius and $m$ is an integration constant. The
cosmological constant is $\Lambda = -6/L^2$. Form M-theory point of
view,  the  eleven-dimensional spacetime  eq.(\ref{ds4}) can be
interpreted  as the near horizon geometry of $N$ coincident
configurations of  M2-branes. In  this background, the AdS radius
$L$ is linked to the
 M2-brane number $N$   via the relation  \cite{a,a7}
\begin{equation}
L^9={N^3/2} \frac{\kappa_{11}^2 \sqrt{2}}{ \pi^5}.
\end{equation}
According to the proposition reported in \cite{Zhang,Zhang1,jhep45},
we consider  the cosmological constant as the number of coincide
M2-branes in the M theory background and its conjugate quantity as
the associated chemical potential.

\begin{figure}[!ht]
\begin{center}
\includegraphics[scale=1.]{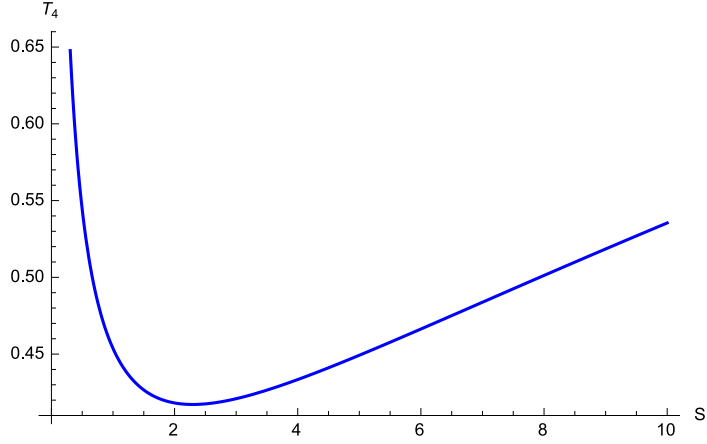}
\end{center}
 \caption{The temperature as function of the entropy $S$, with $N=3$} \vspace*{-.2cm}
 \label{fig1}
\end{figure}

The event horizon $r_h$ of the corresponding  black hole is
determined by solving  the equation $f = 0$.   Exploring
eq.(\ref{f4}), the mass of the black hole can be  written  as
\begin{equation}
M_4=\frac{m \omega_2}{8 \pi  G_4}=\frac{r \omega_2 \left(L^2+r^2\right)}{8 \pi  G_4 L^2}.\footnote{where $\omega_d=\frac{2 \pi ^{\frac{d+1}{2}}}{\Gamma
\left(\frac{d+1}{2}\right)}$.}
\end{equation}
 The Bekenstein-Hawking entropy formula
of the black hole produces
\begin{equation}
S=\left.\frac{A}{4 G_4}\right.=\frac{\omega_2  r^2}{4 G_4}.
\end{equation}
 It is recalled that  four-dimensional Newton gravitational constant is
related to the eleven-dimensional one by
\begin{equation}
G_4=\frac{3 G_{11}
}{2 \pi  \omega_{2}L^{4}}\, 
\end{equation}
For simplicity reason,  we  take in the rest of  the paper
$G_{11}=\kappa_{11}=1$. In this way, the mass of the black hole can
be expressed  as a function of $N$ and $S$
\begin{equation}\label{M4}
M_4(S,N)=\frac{\sqrt{S} \left(16 N+3\ 2^{2/3} \sqrt[3]{\pi } S\right)}{8\ 2^{7/18}
   \sqrt{3} \pi ^{11/18} N^{2/3}}.
\end{equation}
Using  the standard thermodynamic relation $dM = TdS + \mu dN$, the
corresponding  temperature takes the following form
\begin{equation}\label{T4}
T_4=\left.\frac{\partial M_4(S,N)}{\partial S}\right|_N=\frac{8 \sqrt[3]{2} N+9 \sqrt[3]{\pi } S}{8\ 2^{13/18} \sqrt{3} \pi ^{11/18}
   N^{2/3} \sqrt{S}}.
\end{equation}
This quantity  can be identified with  the Hawking temperature of
the black hole.  Using eq. (\ref{M4})The chemical potential $\mu$
conjugate to the number of  M2-branes is given by
\begin{equation}\label{mu4}
\mu_4=\left.\frac{\partial M_4(S,N)}{\partial N}\right|_S=\frac{\sqrt{S} \left(8 N-3\ 2^{2/3} \sqrt[3]{\pi } S\right)}{12\ 2^{7/18}
   \sqrt{3} \pi ^{11/18} N^{5/3}}.
\end{equation}
It defines   the measure of the energy cost to the system when one
increases the variable  $N$.  In terms of these quantities, the
Gibbs free energy  reads as
\begin{equation}\label{gibbs4}
G_4(T,N)=M_4-T_4\  S=\frac{\sqrt{S} \left(8 \sqrt[3]{2} N-3 \sqrt[3]{\pi } S\right)}{8\ 2^{13/18}
   \sqrt{3} \pi ^{11/18} N^{2/3}}.
\end{equation}
Having calculated the relevant  thermodynamical quantities, we
investigate the corresponding  phase transition. To do so, we study
the variation of the Hawking temperature as a function of the
entropy. This variation is plotted in figure \ref{fig1}.

It follows from  figure \ref{fig1} that   the Hawking temperature is
not a monotonic function.  In fact,  it involves a minimum at the
point
\begin{equation}
S_{4,1}=\frac{8}{9} \sqrt[3]{\frac{2}{\pi }} N
\end{equation}
which corresponds to the minimal  temperature
$T_{4_{min}}=\frac{\sqrt{3}}{2 \sqrt[18]{2} \pi ^{4/9}
   \sqrt[6]{N}}$.  It is observed that for such a  temperature no black hole can exist.
    Otherwise, two branches are shown up. Indeed, the first branch associated with small entropy $S$
     values is thermodynamically unstable. However,  the second  phase  corresponding to the large
      entropy $S$ is considered as a thermodynamical stable one.

It is observed  from  Gibbs free energy, given in eq.(\ref{gibbs4}),
that the Hawking-Page phase transition occurs where the
corresponding phase transition temperature is
\begin{equation}
T_{HP_{4}}=\frac{1}{\sqrt[18]{2} \pi ^{4/9} \sqrt[6]{N}}.
\end{equation}
It is verified that this   quantity is larger than the temperature
$T_{4_{min}}=T|_{S=S_{4,1}}$. At the Hawking-Page transition, the
associated entropy takes the following form
\begin{equation}
S_{4,2}=\frac{8}{3} \sqrt[3]{\frac{2}{\pi }} N.
\end{equation}

\begin{figure}[!ht]
\begin{center}
\includegraphics[scale=1.]{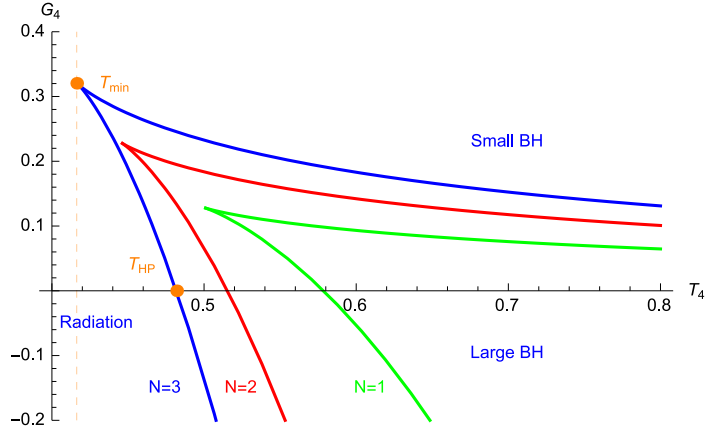}
\end{center}
 \caption{The Gibbs free energy  as function of the Temperature $T_4$, for $N={1,2,3}$.} \vspace*{-.2cm}
 \label{fig2}
\end{figure}

In figure  \ref{fig2}, we illustrate the Gibbs free energy  as
function of  the Hawking temperature $T$ for some fixed values of
$N$.

It is remarked that the  down branch Gibbs free energy for a fixed
$N$ changes its sign at the point $S = S_{4,2}$, which corresponds
to the Hawking-Page transition point.  Moreover, it  is observed  a
minimum temperature  $T_{4_{min}}$  for which no black holes  ($T <
T_{4_{min}}$) can survive.    However, above this temperature, two
branches of the black holes   are shown up. Indeed, the upper branch
describes an unstable small (Schwarzschild-like) black hole
associated with a negative specific heat. For $T > T_{4_{min}}$ ,
the black holes, at lower branch,  can be considered as a stable
solution   corresponding to positive specific heat.
 Since the Hawking-Page temperature $T_{4_{HP}}$ is associated
with vanishing values of the Gibbs free energy,  the black hole
Gibbs free energy becomes negative for $T > T_{4_{HP}}$.   As
reported in \cite{30,15Dolan,Dolan}, at $T = T_{4_{HP}}$,  a  first
order Hawking-Page phase transition occurs
 between the thermal radiations and large black holes.

\begin{figure}[!ht]
\begin{center}
\includegraphics[scale=1.]{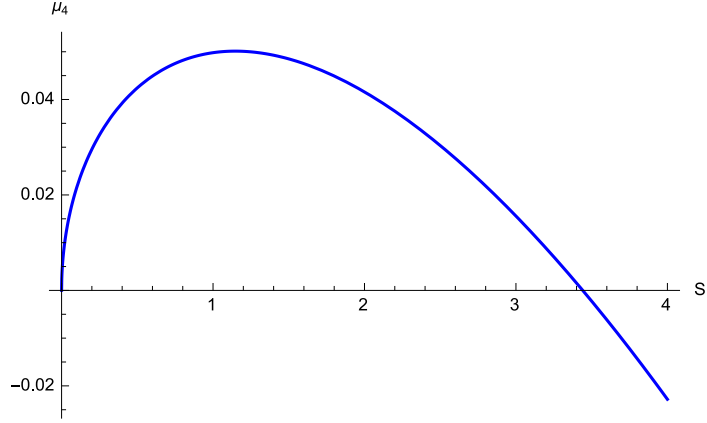}
\end{center}
 \caption{The chemical potential $\mu$ as function of the entropy for  $N=3$} \vspace*{-.2cm}
 \label{fig3}
\end{figure}

To study the phase transition,  we vary    the chemical potential in
terms of the entropy. In figure \ref{fig3},  we plot  such a
variation for  a   fixed value  of  $N$.

For small values of $S$,  the chemical potential is positive.
However, it changes to be negative when $S$ is large. Moreover, the
chemical potential changes its sign at
\begin{equation}
S_{4,3}=\frac{4}{3} \sqrt[3]{\frac{2}{\pi }} N.
\end{equation}
It is easy to cheek the following constraint
\begin{equation}S_{4,3}<S_{4,2}<S_{4,1}.\end{equation}

It turns out that the vanishing of the chemical potential appears in
the unstable branch.

\begin{figure}[!ht]
\begin{center}
\includegraphics[scale=1.]{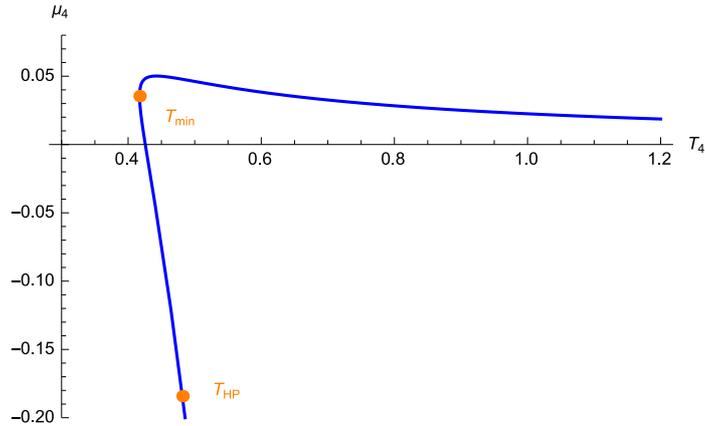}
\end{center}
 \caption{The chemical potential $\mu$ as function of the temperature $T_4$, with $N=3$} \vspace*{-.2cm}
 \label{fig4}
\end{figure}

In figure \ref{fig4}, we plot the chemical potential as a function
of temperature $T_4$
 for a fixed $N$.
 
From figure \ref{fig4}, we can see  the Hawking-Page temperature. On
the branch below this point,  the  black
 holes are stable.  Such a point resides  in the negative region of the chemical potential. However,
a minimum  of the
  temperature the upper branch   which corresponds to unstable black hole solutions   lives  in   the
  positive region of the chemical potential.

\begin{figure}[!ht]
\begin{center}
\includegraphics[scale=1.]{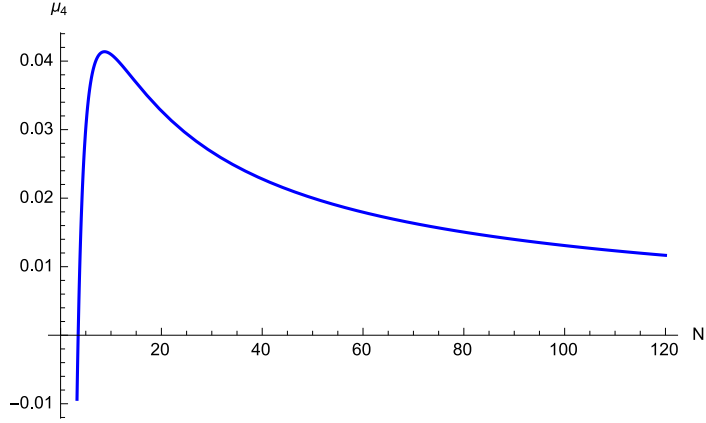}
\end{center}
 \caption{The chemical potential $\mu$ as function of $N$, we have set $S_4=4$.} \vspace*{-.2cm}
 \label{fig5}
\end{figure}

To see the effect  the number of the M2-branes, we discuss the
behavior of the chemical potential $\mu$ in terms of such a
variable. The calculation is illustrated in figure \ref{fig5}.

 It is observed that the chemical potential $\mu$ presents a maximum at
\begin{equation}
N_{4_{max}}=\frac{15}{8} \sqrt[3]{\frac{\pi }{2}} S,\ namely\;\;\ S_{4,4}=\frac{8}{15} \sqrt[3]{\frac{2}{\pi }} N
\end{equation}
It is noted  that $S_{4,4}$ is also less than $S_{4,1}$. It is
remarked that  this  is quite different from the classical gas
having a  negative chemical potential. In the case where  the
chemical potential approaches to zero and becomes positive, quantum
effects should be considered and be relevant in the discussion
\cite{jhep45}.

 Having discussed the case of M2-branes, let us  move  a  higher
 dimensional case provided by M-theory. It is shown that in eleven
 dimensions the dual magnetic of M2-branes  are M5-branes. In the
 following,  we investigate the black holes in such magnetic brane  backgrounds.

 \section{ Thermodynamics of black holes in  $AdS_7\times S^4$ space}

In this section, we discuss the magnetic solution associated with
the near   geometry  $AdS_7\times S^4$.  According to \cite{a58,a},
the corresponding  metric  takes the following form
\begin{equation}\label{ds7}
ds^2= \frac{r}{L}\left(-f dt ^2+\sum_{i=1}^5 dx_i^2 \right)+\frac{L^2}{r^2}f^{-1}dr^2+ L^2 d\Omega_4^2,
\end{equation}
where $d\Omega_4^2$ is the metric of four-dimensional sphere with
unit radius. As in  the case of M2-branes,  the  metric function
reads as

\begin{equation}\label{f7}
f=1-\frac{m}{r^4}+\frac{r^2}{L^2}.
\end{equation}
In M-theory, the  eleven-dimensional spacetime eq.(\ref{ds7}) can be
considered  as the near horizon geometry of $N$ coincident configurations of   M5-branes.
For this solution, the AdS radius $L$  can be related to the number
$N$  via the relation \cite{a,a7}
\begin{equation}
L^9=N^3 \frac{\kappa_{11}^2}{2^7 \pi^5}.
\end{equation}
The mass of the black hole can be computed using eq.(\ref{f7}). The
calculation gives the following expression
\begin{equation}
M_7=\frac{5\  m \omega_5}{8 \pi  G_7}=\frac{5\ r^4 \omega_5 \left(L^2+r^2\right)}{16 \pi  G_7 L^2}.
\end{equation}

\begin{figure}[!ht]
\begin{center}
\includegraphics[scale=1.]{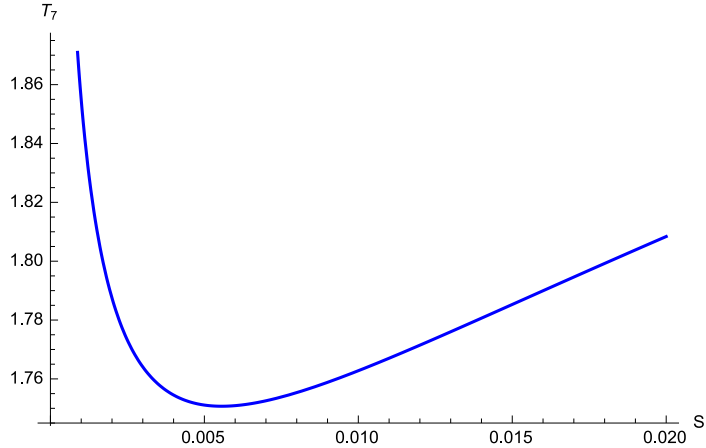}
\end{center}
 \caption{The temperature as function of the entropy, with $N=3$.} \vspace*{-.2cm}
 \label{fig6}
\end{figure}

It is found that the entropy is

\begin{equation}
S=\frac{A}{4 G_7}=\frac{\omega_5  r^5}{4 G_7},\qquad\ \  G_7=\frac{6
G_{11} }{2 \pi  \omega_{5} L^{7}}.
\end{equation}
 Combining these expressions, one can write   the mass in terms of the entropy $S$ and $N$
as follows
\begin{equation}\label{M7}
M_7(S,N)=\frac{5 \left(2^{23/45} 3^{4/5} \pi ^{2/15} N^{8/5}
   S^{4/5}+96\ 2^{2/45} \sqrt[5]{3} S^{6/5}\right)}{48 \pi
   ^{23/45} N^{17/15}}.
\end{equation}
The Hawking temperature can be obtained using the first law  of
thermodynamics $dM = TdS + \mu dN$. Indeed, it is given by

\begin{equation}\label{T7}
T_7=\left.\frac{\partial M_7(S,N)}{\partial S}\right|_N=\frac{2^{23/45} 3^{4/5} \pi ^{2/15} N^{8/5}+144\
   2^{2/45} \sqrt[5]{3} S^{2/5}}{12 \pi ^{23/45}
   N^{17/15} \sqrt[5]{S}}.
\end{equation}
It is found,  after calculations, that the chemical potential $\mu$,
conjugate to the number of M5-branes,

\begin{equation}\label{mu7}
\mu_7=\left.\frac{\partial M_7(S,N)}{\partial N}\right|_S=\frac{7\ 2^{23/45} 3^{4/5} \pi ^{2/15} N^{8/5}
   S^{4/5}-1632\ 2^{2/45} \sqrt[5]{3} S^{6/5}}{144 \pi
   ^{23/45} N^{32/15}}.
\end{equation}
 Similarly, the Gibbs free energy can be computed.   It is given by

\begin{equation}\label{gibbs7}
G_7(T,N)=M_7-T_7\  S=\frac{2^{23/45} 3^{4/5} \pi ^{2/15} N^{8/5}
   S^{4/5}-96\ 2^{2/45} \sqrt[5]{3} S^{6/5}}{48 \pi ^{23/45}
   N^{17/15}}
\end{equation}
As in the previous case, the stability discussion can be done by
varying the two variables $S$ and $N$. We first deal with the phases
transition. Indeed, it  can be studied in terms of monotony of the
Hawking temperature in terms of the entropy. This variation is
plotted in figure \ref{fig6}.

 We can clearly  see that the Hawking temperature is not a monotonic function. It involves a minimum at the point
\begin{equation}
S_{7,1}=\frac{\sqrt[3]{\pi } N^4}{6912\ 2^{5/6} \sqrt{3}}
\end{equation}
associated with  the temperature $T_{7_{min}}=\frac{2\ 2^{5/18}
\sqrt{3}}{\pi ^{4/9} \sqrt[3]{N}}$.  It is observed that for the
minimal temperature no black hole can survive. Otherwise, two
branches appear. Indeed, the first branch associated with small
entropy values is thermodynamically unstable.  The second one
corresponding to the large entropy is considered as a
thermodynamically stable branch.

It follows from the Gibbs free energy, given in eq.(\ref{gibbs7}),
that the Hawking-Page phase transition occurs where the
corresponding phase transition temperature
\begin{equation}
T_{HP_{7}}=\frac{2\ 2^{5/18} \sqrt{3}}{\pi ^{4/9} \sqrt[3]{N}}
\end{equation}

This quantity is larger than the temperature $T_{7_{min}}=T|_{S=S_{7,1}}$. At the Hawking-Page transition, the
corresponding entropy is given by
\begin{equation}
S_{7,2}=\frac{\sqrt[3]{\frac{\pi }{2}} N^4}{6144}
\end{equation}

\begin{figure}[!ht]
\begin{center}
\includegraphics[scale=1.]{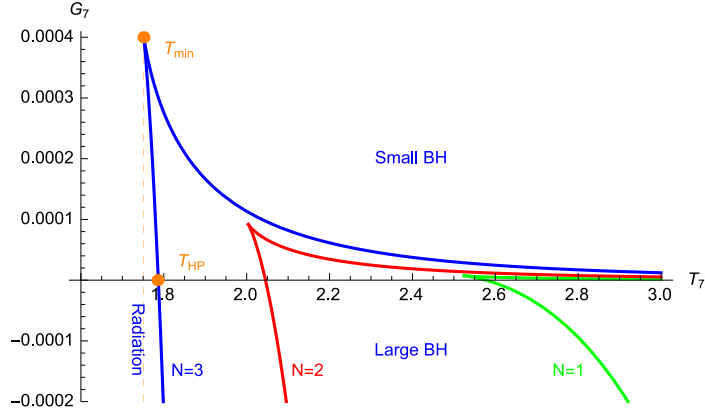}
\end{center}
 \caption{The Gibbs free energy  as function of the Temperature, for $N={1,2,3}$.} \vspace*{-.2cm}
 \label{fig7}
\end{figure}

The figure \ref{fig7} illustrates the Gibbs free energy with respect
to the Hawking Temperature $T$ for some fixed  values  of $N$.

 For a fixed $N$, it  follows that the down branch Gibbs free energy   changes its sign
at the point $S = S_{7,2}$, corresponding   to the Hawking-Page
transition point.

\begin{figure}[!ht]
\begin{center}
\includegraphics[scale=1.]{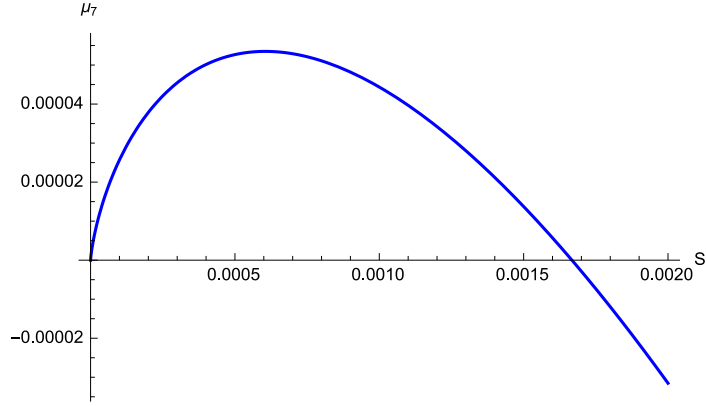}
\end{center}
 \caption{The chemical potential $\mu$ as function of the entropy, with $N=3$} \vspace*{-.2cm}
 \label{fig8}
\end{figure}

To study  the phase transition, we vary the chemical potential in
terms of  the entropy. In Fig. \ref{fig8},  we plot  such a
variation  by fixing  the  number of M5-branes in M-theory.

We see that the chemical potential is positive when we consider
small values of  $S$. However, it changes to be negative  for large
values of  $S$. The chemical potential changes its sign at
\begin{equation}
S_{7,3}=\frac{49 \sqrt{\frac{7}{17}} \sqrt[3]{\frac{\pi }{2}}
   N^4}{1775616}.
\end{equation}
As in the case of M2-branes,  one has
\begin{equation}S_{7,3}<S_{7,2}<S_{7,1}.\end{equation}

As we will see that the vanishing of the chemical potential appears
in the unstable branch. Similar behaviors appeared in the case of
M2-branes.  This implies that the vanishing of the chemical
potential does not make any sense from the point of view of dual
conformal field theory. This point deserves a deeper study. We hope
to comeback to this point in future.

\begin{figure}[!ht]
\begin{center}
\includegraphics[scale=1.]{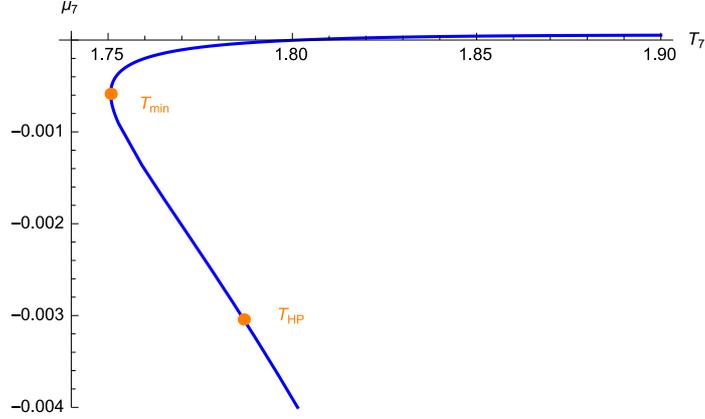}
\end{center}
 \caption{The chemical potential $\mu$ as function of the temperature $T_7$, with $N=3$} \vspace*{-.2cm}
 \label{fig9}
\end{figure}

To see the effect of the temperature,  figure \ref{fig9}  presents
the chemical potential as a function of  the temperature $T_7$ for
fixed values of  $N$. It is noted that one has  similar behaviors
appeared in the case of M2-branes.

\begin{figure}[!ht]
\begin{center}
\includegraphics[scale=1.]{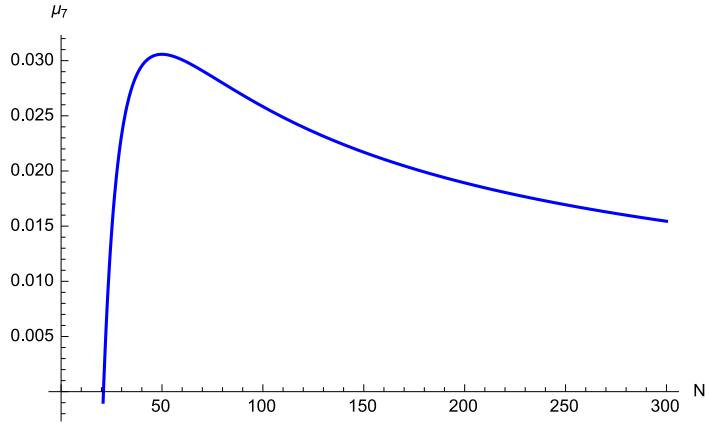}
\end{center}
 \caption{The chemical potential $\mu$ as function of $N$, we have set $S_7=4$.} \vspace*{-.2cm}
 \label{fig10}
\end{figure}

 Figure \ref{fig10} shows  the chemical potential  as a
function of $N$ in the case with a fixed entropy.

 We can observe that the chemical potential $\mu$ presents a maximum at
\begin{equation}
N= 16 \left(\frac{17}{7}\right)^{5/8} \sqrt[4]{3} \sqrt[12]{\frac{2}{\pi }} \sqrt[4]{S},\ namely\;\;                          S_{7,4}=\frac{49 \sqrt{\frac{7}{17}} \sqrt[3]{\frac{\pi }{2}} N^4}{56819712}.
\end{equation}

In the following section, we will study thermodynamical geometry
 of the M2 and M5-branes  black holes in the extended phase space
 to reconsider the study of the stability  problem.

\section{Geothermodynamics of AdS  black holes in M-theory}
In this section, we discuss the  geothermodynamics AdS black holes
in $AdS_{p+2}\times S^{11-p-2}$. This study concerns singular limits
of certain thermodynamical quantities including the  heat capacity.
This quantity is the relevant in the study of  the stability of such
black hole solutions.

\begin{center}
\begin{figure}[!ht]
\begin{tabbing}
\hspace{9cm}\=\kill
\includegraphics[scale=.9]{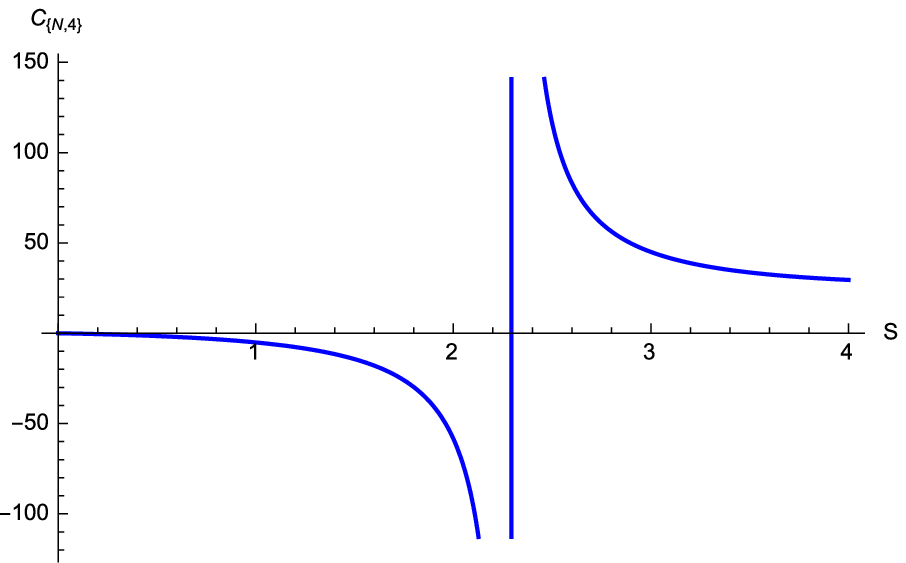}\>\includegraphics[scale=.9]{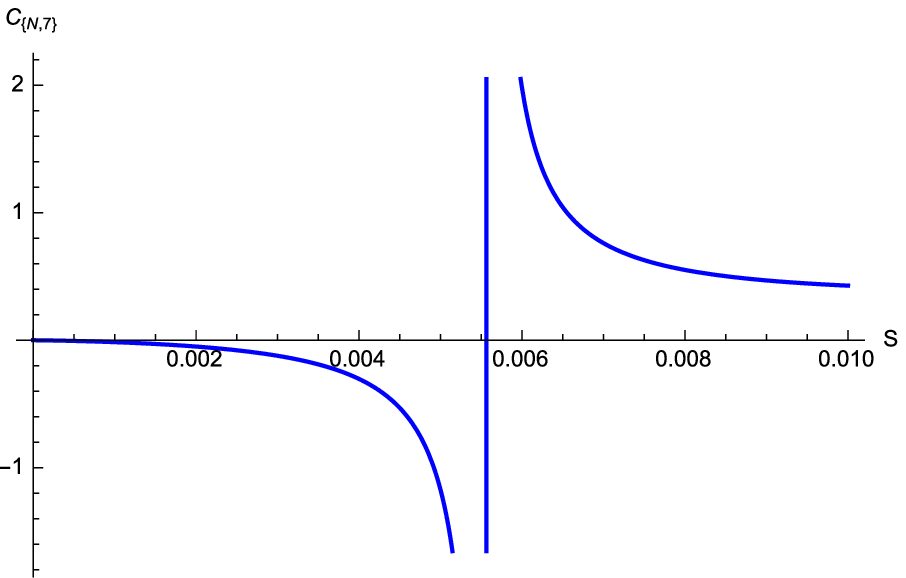} \\
\end{tabbing}
\vspace*{-.2cm} \caption{ The heat capacity in the case with a fixed $N = 3$ as a function of entropy $S$ for the two backgrounds.}
\label{figcn}
\end{figure}
\end{center}

To elaborate this discussion, the  number of branes $N$ should be
fix  to consider a  canonical ensemble. For a fixed $N$, the heat
capacity  for M2 and M5-branes   are  given respectively by

\begin{eqnarray}
C_{N,4}&=& T_4\left(\frac{\partial S}{\partial T_4}\right)_{N}=\left(\frac{1}{\frac{8}{9} \sqrt[3]{\frac{2}{\pi }}
   N+S}-\frac{1}{2 S}\right)^{-1}\\
C_{N,7}&=& T_7\left(\frac{\partial S}{\partial T_7}\right)_{N}=\frac{720 S^{7/5}+5\ 2^{7/15} 3^{3/5} \pi ^{2/15}
   N^{8/5} S}{144 S^{2/5}-2^{7/15} 3^{3/5} \pi
   ^{2/15} N^{8/5}}.
\end{eqnarray}
These equations  contain many interesting  thermodynamical
properties. Indeed,   the heat capacity involves a divergence   at
the point of $S_{i,1_{i\in\{4,7\}}}$. For a fixed $N$, this point
can be identified
   with the point corresponding to the minimal
   Hawking temperature.   In the case  $S <S_{i,1_{i\in\{4,7\}}}$,
    the heat capacity is negative  showing  the thermodynamical
    instability. However, it becomes
     positive  in the region defined by   $S > S_{i,1_{i\in\{4,7\}}}$.  These  behavior of
     $C_{N,i_{i\in\{4,7\}}}$ as a function of $S$  can be illustrated in  figure\ref{figcn}.

  To show the  singularity of   the corresponding   heat
 capacity, the thermodynamical geometry of    such black hole
 solutions should be discussed including the thermodynamical
 curvature. To compute such a quantity, one can use several metrics.
 However, we can explore the Quevedo  metric  which reads as \cite{hep54,hep55,hep56,hep57}

\begin{equation}
g^Q=(S T+N\mu) \left(\begin{array}{cc}M_{S\ S} & 0 \\0& M_{N\ N}\end{array}\right),
\end{equation}
where $M_{i\ j}$
stands for $\partial^2 M/ \partial x^i\partial x^j$, and $x^1=S$, $x^2=N$.
 The scalar curvature of this metric can be computed in a direct
 way.  For M2 and M5-branes respectively, the calculation  gives the
 following expressions
\begin{equation}
R_4^Q=\frac{55296 \sqrt[9]{2} \pi ^{14/9} N^{7/3} \left(-8192 N^3-729 \pi  S^3+9792 \sqrt[3]{2} \pi ^{2/3} N
   S^2+2112\ 2^{2/3} \sqrt[3]{\pi } N^2 S\right)}{5 \left(15\ 2^{2/3} \sqrt[3]{\pi } S-16 N\right)^2 \left(9 \sqrt[3]{\pi
   } S-8 \sqrt[3]{2} N\right)^2 \left(8 \sqrt[3]{2} N+3 \sqrt[3]{\pi } S\right)^3}
\end{equation}
and
\begin{equation}
R_7^Q=\frac{A}{B \ C}.
\end{equation}
The quantities $A$, $B$  and $C$ are given by
\begin{eqnarray}
A&=&414720\ 2^{2/45} \pi ^{52/45} N^{58/15} \left(361361\ 2^{7/15} \sqrt[5]{3} \pi
   ^{8/15} N^{32/5}+82685568\ 3^{3/5} \pi ^{2/5} N^{24/5}
   S^{2/5}\right.\\ \nonumber
   &+&\left.388512000\ 2^{8/15} \pi ^{4/15} N^{16/5} S^{4/5}+6806372352
   \sqrt[15]{2} 3^{2/5} \pi ^{2/15} N^{8/5} S^{6/5}+5474746368\ 2^{3/5} 3^{4/5}
   S^{8/5}\right)\\
   B&=& S^{6/5} \left(-133\ 3^{3/5} \pi ^{4/15} N^{16/5}+61680\ 2^{8/15} \pi ^{2/15}
   N^{8/5} S^{2/5}+104448 \sqrt[15]{2} 3^{2/5} S^{4/5}\right)^2\\
   C&=&\left(-19\ 3^{3/5} \pi ^{4/15} N^{16/5}+1320\ 2^{8/15} \pi ^{2/15}
   N^{8/5} S^{2/5}+2304 \sqrt[15]{2} 3^{2/5} S^{4/5}\right)^2
\end{eqnarray}

\begin{center}
\begin{figure}[!ht]
\begin{tabbing}
\hspace{9cm}\=\kill
\includegraphics[scale=1]{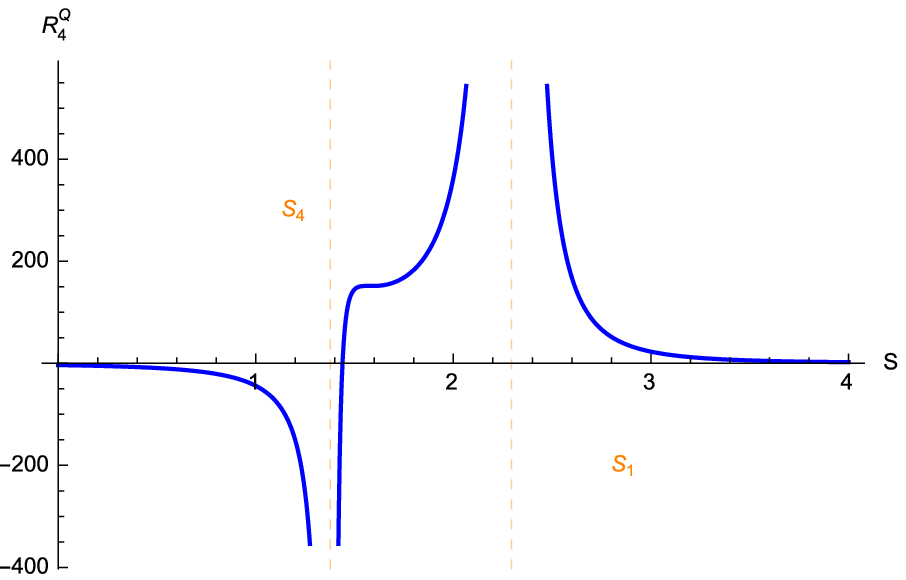}\>\includegraphics[scale=1]{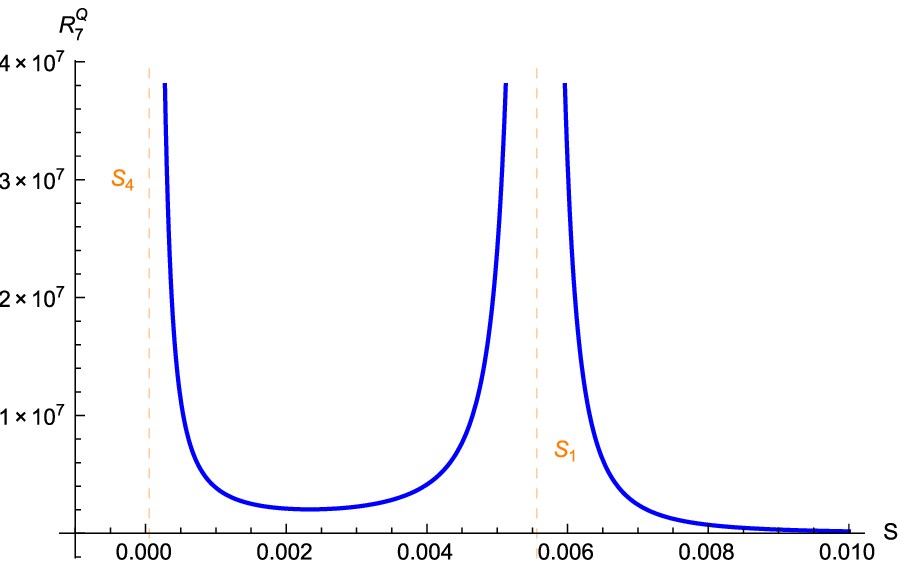} \\
\end{tabbing}
\vspace*{-.2cm} \caption{The scalar curvature vs entropy for the Quevedo metric case with $N = 3$  for different backgrounds}
\label{figrq}
\end{figure}
\end{center}

In  figure \ref{figrq}, we plot the scalar curvature of the Quevedo
metric as a function of  the entropy for M2 and M5-branes.

It follows  from  figure \ref{figrq}   that one has  two  divergent
points located at $S_{i,4_{i\in\{4,7\}}}$  and
$S_{i,1_{i\in\{4,7\}}}$ respectively. The first one  coincides with
the divergent point of $C_{N}$. However,  the second one is
associated with the  maximum of  the chemical potential considered
as a function of $N$. It is worth noting that this result is in good
agrement with  the recent study  reported  in \cite{67, 68,hep65}
saying  that the divergences of scalar curvature for the Quevedo
metric corresponds to the divergence or zero for the  heat capacity.
It has been suggested that these results can be explored to
understand the  link between  the phase transition and the
thermodynamical curvature.

\section{Conclusion}

In this paper, we have   investigated  the thermodynamics and
thermodynamical geometry of AdS black holes from M2 and M5-branes.
Concretely, we have considered    AdS black holes in
$AdS_{p+2}\times S^{11-p-2}$, where $p=2,5$ by viewing the number of
M2 and M5-branes as a thermodynamical variable. First, we   have
discussed the corresponding phase transition by computing the
relevant quantities. For M2 and M5-branes, we have computed the
chemical potential and discussed the corresponding stabilities.
 Then,  we have studied  the thermodynamical geometry
associated with   such  AdS black holes.  More precisely,
 we have computed the  scalar curvatures  from the  Quevedo metric.  The calculations  show
 similar thermodynamical properties  appearing in  the phase transition program.  This present work, concerning M-theory,  may support
 the relation  between   the  phase transition
and divergence of thermodynamical curvature  studied in type IIB
superstring.

This work poses a  question concerning a  9-dimensional AdS black
holes  associated with D7-branes on $AdS_9$-space. In fact, it may
be possible to consider  a geometry of the form $$AdS_{9}\times
S^{1}\times T^{2}$$ inspired by the recent work on black holes in
F-theory\cite{vafa}. This   may support the results concerning the
link between  the phase transition and the thermodynamical
curvature.

\end{document}